\documentclass[conference]{IEEEtran}
\IEEEoverridecommandlockouts

\usepackage{cite}
\usepackage{amsmath,amssymb,amsfonts}
\usepackage{algorithmic}
\usepackage{graphicx}
\usepackage{textcomp}
\usepackage{xcolor}
\def\BibTeX{{\rm B\kern-.05em{\sc i\kern-.025em b}\kern-.08em
    T\kern-.1667em\lower.7ex\hbox{E}\kern-.125emX}}

\newcommand{\ignore}[1]{}

\usepackage{mathtools}
\newcommand{\ours}{TACO\xspace}
\usepackage{lipsum}
\usepackage{xspace}
\usepackage{booktabs}
\usepackage{tabularx}
\usepackage{makecell}
\usepackage{braket}
\usepackage{enumitem}
\usepackage{balance}
\usepackage{xcolor}

\newcommand{\mstrut}{\rule{0pt}{2.4ex}} %
\usepackage{multirow}

\usepackage[
  colorlinks=true,
  linkcolor=blue,
  citecolor=blue,
  urlcolor=blue
]{hyperref}

\begin{document}

\title{Transpiler-Architecture Co-Design to Curb Clifford Costs in Fault-Tolerant Quantum Computing}

\author{
\IEEEauthorblockN{
Meng Wang\textsuperscript{1, 2},
Chenxu Liu\textsuperscript{2},
Samuel Stein\textsuperscript{2},
Yufei Ding\textsuperscript{3},
Poulami Das\textsuperscript{4},
Prashant J. Nair\textsuperscript{1},
Ang Li\textsuperscript{2,5}
}

\IEEEauthorblockA{\textsuperscript{1}
\textit{Department of Electrical and Computer Engineering, The University of British Columbia}}

\IEEEauthorblockA{\textsuperscript{2}
\textit{Physical and Computational Science Division, Pacific Northwest National Laboratory}}

\IEEEauthorblockA{\textsuperscript{3}
\textit{Department of Computer Science and Engineering, University of California San Diego}}

\IEEEauthorblockA{\textsuperscript{4}
\textit{Department of Electrical and Computer Engineering, The University of Texas at Austin}}

\IEEEauthorblockA{\textsuperscript{5}
\textit{Department of Electrical and Computer Engineering, University of Washington}}
}

\maketitle

\thispagestyle{plain}
\pagestyle{plain}

\begin{abstract}
Quantum Error Correction (QEC) codes form the foundation of Fault-Tolerant Quantum Computing (FTQC) and predominantly use the Clifford+T gate set. Recently, Clifford operations have become the key performance bottleneck in implementing QEC. While state-of-the-art approaches like Pauli-Based Compilation (PBC) reduce Clifford overhead by transforming Clifford gates into Pauli measurements, they do so at the cost of gate-level parallelism, inflating circuit depth and execution times.

To overcome these limitations, we introduce \ours, a \textit{T}ranspiler–\textit{A}rchitecture \textit{Co}-design framework that tackles the Clifford bottleneck through circuit and architectural optimization. \ours uses FTQC insights to guide hardware-aware Clifford gate elimination and circuit restructuring, and leverages the resulting optimized circuits to refine architectural design. \ours applies FTQC-specific transformations to aggressively reduce Clifford overhead from rotation synthesis and Toffoli decompositions, while preserving gate-level parallelism. The resulting architecture is optimized for the locality and data-movement patterns of these circuits, enabling high-throughput, resource-efficient execution. Our evaluation across diverse benchmarks shows that \ours achieves up to \textbf{21.9$\times$} (mean \textbf{4.4$\times$}) reduction in execution time compared to the state-of-the-art baseline.
\end{abstract}


\section{Introduction}
Quantum Error Correction (QEC)~\cite{shor1995scheme, calderbank1996good, gottesman1997stabilizer, knill2001scheme, kitaev2003fault} is essential to enable practical fault-tolerant quantum computing (FTQC). QEC encodes logical qubits into multiple physical qubits to actively detect and correct errors~\cite{gambetta2017building,google2023suppressing,bluvstein2024logical,reichardt2024logicalcomputationdemonstratedneutral}. Among QEC codes, the surface code~\cite{kitaev2003fault, fowler2012surface, horsman2012surface} is especially promising, offering planar connectivity, high thresholds, and efficient decoding, and can realize universal fault-tolerant computation through additional mechanisms such as lattice surgery, code deformation, and magic-state distillation~\cite{bravyi2005universal, fowler2012surface, nielsen2010quantum}. As quantum algorithms advance, balancing the resource costs of Clifford and T gates in QEC is key to realizing practical FTQC.

\begin{figure}[t]
    \centering
    \includegraphics[width=\linewidth]{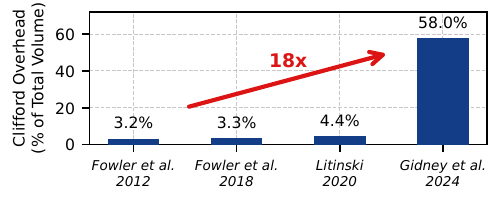}
    \caption{Clifford gate volume percentage of the 18-qubit QFT circuit execution increased by more than 18$\times$ from 3.2\% to 58\% as T gates become cheaper with improved magic state preparation protocols~\cite{fowler2012surface, fowler2018low, litinski2019magic, gidney2024magic}.}
    \label{fig:t_overhead}
\end{figure}

Historically, magic-state preparation for T gates dominated FTQC resource costs, often exceeding 95\% of the total qubit-cycle volume. This challenge motivated substantial progress in T-gate optimization, significantly reducing T-state overheads~\cite{Campbell_2017, haah2017magic, gidney2019efficient, litinski2019magic, gidney2024magic}. As a result, FTQC resource estimates have begun to expose other costs that were previously treated as negligible. In particular, Clifford operations can now constitute a major, and in some cases dominant, fraction of the computational overhead. As shown in Figure~\ref{fig:t_overhead}, the Clifford-overhead fraction in an 18-qubit QFT circuit increases from 3.2\% to 58\% under resource estimates spanning the past decade. We observe a similar trend across a broad set of benchmark circuits, as discussed in Section~\ref{sec:motivation_clifford_bottleneck}.

\begin{figure*}
    \centering
    \includegraphics[width=\linewidth]{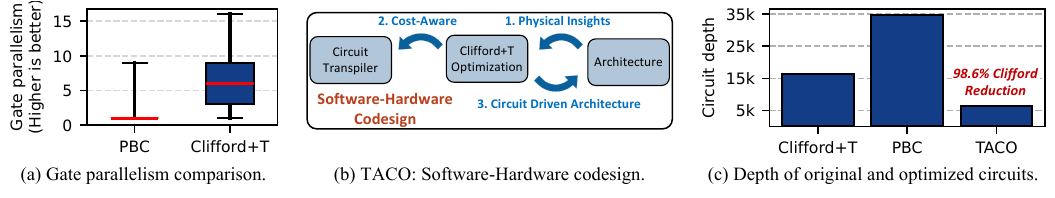}
    \caption{Comparison of Clifford optimization approaches showing gate reduction vs. circuit depth trade-offs for an 18-qubit QFT circuit. (a) The state-of-the-art Clifford optimization method, PBC, significantly restricts gate parallelism. (b) \ours employs a software-hardware co-design approach that (c) reduces Clifford gates by 98.6\% in the QFT circuit, which is comparable to PBC, while achieving 5.3$\times$ lower circuit depth.}
    \label{fig:framework_preview}
\end{figure*}

\noindent\textbf{Limitations of State-of-the-Art:} A common approach for addressing Clifford gate overhead is to convert Clifford+T circuits into Pauli-based circuits (PBC)~\cite{Bravyi2016PBC, litinski2019game}. This transformation commutes all Clifford gates to the end of the circuit, effectively absorbing them into the final measurements and eliminating explicit Clifford operations. However, this approach comes with important trade-offs. PBC introduces numerous multi-qubit $\frac{\pi}{4}$ rotations. These rotations significantly constrain gate parallelism (i.e., the number of gates that can be executed in parallel), as shown in Figure\ref{fig:framework_preview} (a). Moreover, as multi-qubit logical operations involve more qubits, their error rates rise, often requiring higher code distances to maintain fault tolerance. This, in turn, increases physical qubit requirements and further limits the efficiency of PBC.

\vspace{0.05in}
\noindent\textbf{Our Proposal:} We address the Clifford bottleneck with \ours, a systematic \textbf{T}ranspiler–\textbf{A}rchitecture \textbf{C}o-design \textbf{O}ptimization framework. We observe that optimizing either the circuit or hardware in isolation cannot efficiently resolve the trade-offs between gate count, parallelism, and hardware cost in FTQC. \ours overcomes these challenges through three tightly connected strategies: (1) \textit{Hardware-informed Clifford optimization} removes Clifford gates by analyzing their origins and exploiting hardware-native operations. This reduces Clifford count by over 91\% on average while maintaining parallelism. (2) \textit{FTQC-aware transpilation} leverages the simplified Clifford structure to map circuits into fault-tolerant forms that align with the true hardware cost hierarchy. This ensures both Clifford and T gates are synthesized efficiently. (3) \textit{Resource-locality-driven architecture} exploits the locality and regularity in optimized circuits to allocate compute and storage resources efficiently. This helps sustain high gate throughput and minimizes physical qubit overhead.

\noindent\textbf{Insight 1: Hardware-Informed Clifford+T Optimization:} We use the insight that Clifford gates in FTQC circuits mainly arise from two sources: the decomposition of Toffoli gates and the synthesis of single-qubit rotations such as $R_z$. In typical quantum algorithms, each Toffoli gate decomposes into 8 Clifford and 7 $T$ gates. We show that these Clifford gates can be merged or canceled \emph{locally} based on algebraic structure, without reducing gate parallelism. For single-qubit rotations, standard Clifford+$T$ synthesis can expand each rotation into hundreds of gates, with Cliffords comprising up to 60\% of the sequence. Unlike the Toffoli case, these Clifford gates are not locally removable within Clifford+$T$, since they serve as basis changes between $T$ gates ($R_z(\frac{\pi}{4})$). However, recognizing that $R_x(\frac{\pi}{4})$ can be implemented using mechanisms analogous to $T$ gates expands the effective gate set, making most basis changes unnecessary and reducing Clifford gates by over 98\%. By combining local Clifford gate cancellation for Toffoli decompositions with a hardware-native implementation of single-qubit rotations, we eliminate over 98.6\% of Clifford gates. As shown in Figure~\ref{fig:framework_preview}(c), our optimized Clifford+T circuits achieve a 5.3$\times$ lower circuit depth compared to PBC.

\noindent\textbf{Insight 2: FTQC-Aware Circuit Transpilation:} Transpiling algorithmic circuits into Clifford+$T$ form involves two key steps: decomposition and gate synthesis. Although decomposing circuits into standard gate sets may appear straightforward, we find that it offers significant, often hidden optimization opportunities that are largely missed by existing transpilers~\cite{vartiainen2004efficient,nielsen2010quantum,bergholm2018pennylane,qiskit2024}. Our FTQC-aware transpiler exploits these opportunities by jointly coordinating decomposition and synthesis to minimize the resulting Clifford+$T$ gate count.

\noindent\textbf{Insight 3: Resource-Locality-Driven Architecture:}
We exploit the high spatial and temporal locality of our optimized FTQC circuits, especially in extended $\frac{\pi}{4}$ rotation sequences, to design an architecture with dedicated compute and memory blocks. Compute qubits, characterized by frequent gate activity, are clustered in compute blocks optimized for rapid access to magic state resources and efficient execution of consecutive $\frac{\pi}{4}$ rotations. These blocks are configured to expose both the $X$ and $Z$ edges of each logical qubit, enabling flexible, low-latency lattice surgery with state-distillation ancillae. Storage qubits, which remain idle except during Clifford operations, are assigned to memory blocks arranged in a compact three-row layout. This layout minimizes physical area while preserving connectivity via shared ancilla tiles. This enables rapid (typically within a few QEC cycles) and flexible transfer of logical qubits between memory and compute blocks.

At a higher level, our approach systematically integrates hardware layout, circuit transpilation, and architectural refinement through a cross-layer feedback loop. Hardware layout decisions, such as compute and memory block partitioning and ancilla tile placement, directly inform transpiler optimizations. This allows Clifford+T circuits to be mapped to maximize locality and throughput. Conversely, insights from circuit transpilation, such as $\frac{\pi}{4}$ rotation clustering and gate reuse, drive targeted refinements in the hardware design. This iterative co-design allows each layer to reinforce the others, resulting in a 2.3$\times$ reduction in the total Clifford+T gate count. From the resulting circuit, we further eliminate, on average, 91\% of remaining Clifford gates while preserving parallelism, leading to a geometric mean of 4.4$\times$ speedup in QEC cycles. Our resource-locality-driven architecture sustains one logical gate per QEC cycle with just $1.5n + 4$ logical qubit tiles. This approach significantly outperforms prior designs that require $2n+\sqrt{8n}+1$ logical qubit tiles~\cite{litinski2019game}.

\begin{figure*}
    \centering
    \includegraphics[width=\linewidth]{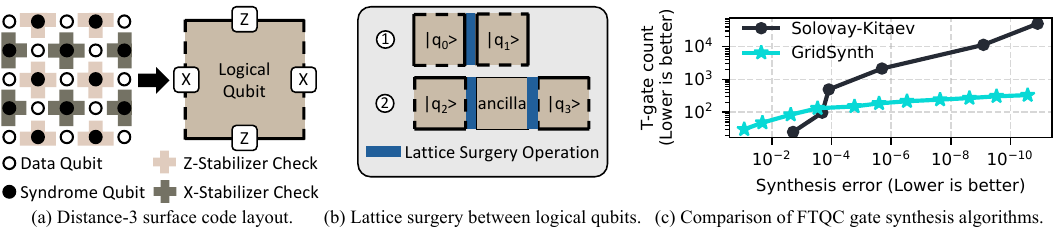}
    \caption{(a) Distance-3 surface code layout showing data qubits ($\circ$), syndrome qubits ($\bullet$), stabilizers (shaded areas), and logical qubit abstraction. X and Z-stabilizers are shown with different colors. (b) Lattice surgery operations between \textcircled{1} neighboring logical qubits and \textcircled{2} non-neighboring logical qubits using an ancilla logical qubit. (c) T-gate count versus synthesis error when synthesizing a random 1-qubit unitary into a Clifford+T gate sequence using the Solovay-Kitaev algorithm~\cite{kitaev1997quantum, dawson2005solovay} and GridSynth~\cite{ross2014optimal}.}
    \label{fig:background}
\end{figure*}

\vspace{0.1in}
\noindent In summary, this work has \emph{four} key contributions:

\begin{itemize}[leftmargin=0cm,itemindent=.5cm,labelwidth=\itemindent,labelsep=0cm,align=left, listparindent=0.1cm,itemsep=0in]
    \item We identify Clifford gates as the dominant, yet previously overlooked source of overhead in FTQC, contributing to more than half of the total gate overhead. 
    \item We propose \ours, a cross-stack framework that reduces Clifford gate overhead by 91\% on average (up to 98.6\%). This leads to a mean 4.4$\times$ speedup across diverse benchmarks.
    \item We develop an FTQC-aware transpiler that reduces overall gate count by 2.3$\times$ compared to state-of-the-art compilers.
    \item We design a hardware architecture tailored to the structure and locality of optimized circuits, achieving one logical gate per QEC cycle with only $1.5n{+}4$ logical qubit tiles. This approach substantially outperforms prior designs that require $2n{+}\sqrt{8n}{+}1$ logical qubit tiles.

\end{itemize}

\section{Background}
\label{sec:background}
\subsection{Surface Code and Fault-Tolerant Operations}

\noindent\textbf{Basics of Surface Code:} Quantum error correction (QEC) protects fragile qubits by encoding logical qubits into multiple physical qubits. The surface code~\cite{bravyi1998quantum, dennis2002topological, fowler2012surface} arranges qubits in a 2D lattice with data qubits ($\circ$) and syndrome qubits ($\bullet$) that perform stabilizer measurements to detect errors (Figure~\ref{fig:background}(a)). Each surface code patch defines a logical qubit with distance $d$, capable of correcting up to $\frac{d-1}{2}$ errors.

\noindent\textbf{Lattice Surgery:} Logical fault-tolerant operations are implemented via \textit{lattice surgery}~\cite{horsman2012surface, fowler2018low, erhard2021entangling, tan2024sat}, which enables qubit interactions by merging or splitting patches along shared edges (Figure~\ref{fig:background}(b)). For non-adjacent qubits, interactions are facilitated by repositioning or using ancillary qubits.

\noindent\textbf{Gate Implementation:} The surface code natively supports most Clifford gates (Pauli-X/Y/Z, Hadamard, CNOT)~\cite{fowler2012surface, litinski2019game}, while the Phase gate (S) requires code deformations or gate teleportation with $\ket{Y}$ states. The non-Clifford T gate, essential for universal computation, is implemented through gate teleportation using magic states $\ket{M}$ prepared via resource-intensive magic state distillation~\cite{bravyi2012magic, jones2013multilevel, litinski2019magic, gidney2024magic}.

\subsection{Qubit-Cycle Cost of FTQC}
\label{sec:background_spact_time_cost}
The overhead of FTQC execution is quantified using the \textit{qubit-cycle volume} metric, where \textit{qubit} denotes physical qubits and \textit{cycle} is measured in QEC cycles. Each cycle represents one round of syndrome measurement and error correction, and the total volume is the product of these quantities. The FTQC circuit size determines the total number of logical qubits. In contrast, the number of physical qubits per logical qubit depends on the chosen code distance ($d$), set by the required logical error rate and total number of cycles.

The circuit's critical path determines the total number of QEC cycles. Within the Clifford+T gate set, the execution cost of each gate type differs. Pauli gates can be applied with no cost~\cite{fowler2012surface}. CNOT gates use lattice surgery and require \textit{3$d$+4} cycles (where d is the code distance). Hadamard gates require patch-deformation techniques, requiring \textit{3$d$+4} cycles. The S gate requires preparing and injecting a $\ket{Y}$ state, with the total cost, including both preparation and consumption, taking \textit{1.5$d$+3} cycles. For the T gate, once a magic state is prepared, consuming it takes \textit{2.5$d$+4} cycles. These cost estimates are based on the recent resource estimation study~\cite{blunt2024compilation}.

\subsection{Fault Tolerant Quantum Computing: Clifford+T Gates}
\label{sec:motivation_circ_synthesis}
To execute an FTQC circuit, all gates must be expressed (transpiled) in the Clifford+T basis. Some gates, such as the Toffoli, can be exactly decomposed into Clifford+T gates. However, many gates, especially single-qubit rotations, do not have an exact representation and instead require approximate synthesis into sequences of Clifford+T gates.

The Solovay-Kitaev algorithm~\cite{kitaev1997quantum, dawson2005solovay} was one of the first methods proposed for this synthesis task, offering asymptotically efficient approximations for arbitrary unitary gates. The general algorithm can accept any unitary and generate approximations in any universal gate set that includes inverses for all its gates. It produces a sequence of length $O\left(\log^c(1/\epsilon)\right)$ to achieve an approximation error $\epsilon$, with practical $c \approx 3.97$~\cite{dawson2005solovay}, where $c$ is a constant. Thus, the resulting gate sequences are prohibitively long even for moderate accuracy.

Recent number-theoretic approaches such as GridSynth~\cite{ross2014optimal} have dramatically improved synthesis efficiency by specifically targeting single-qubit Z-axis rotations (Rz gates). As shown in Figure~\ref{fig:background}(c), GridSynth achieves an error below $10^{-10}$ with just 332 T gates, compared to over 50,000 gates using Solovay-Kitaev. Since any quantum algorithm can be decomposed into Rz plus Clifford gates, GridSynth is adopted as the default synthesis method in our framework, ensuring both accuracy and practical gate counts for large-scale FTQC circuits.

\section{Motivation}
\label{sec:motivation}

\subsection{The Decadal Shift: T-Gate Costs Down, Clifford Costs Up}
\label{sec:motivation_clifford_bottleneck}

\begin{figure}[h!]
    \centering
    \includegraphics[width=\linewidth]{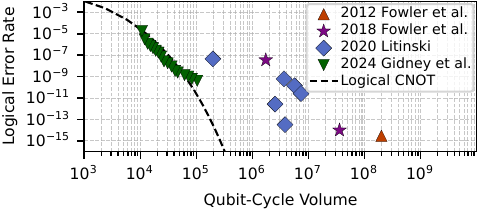}
    \caption{Magic state preparation overhead (qubit-cycle volume) has decreased by more than 100$\times$ since 2012 and is now comparable to logical CNOT operations. Data from~\cite{fowler2012surface, fowler2018low, litinski2019magic, gidney2024magic}.}
    \label{fig:t_cost}
\end{figure}

Historically, T gates were the primary driver of FTQC overhead, with magic state preparation requiring over 1000$\times$ more resources than Clifford operations~\cite{bravyi2005universal,fowler2012surface}. This motivated a decade of advances in T-gate and magic-state optimization~\cite{fowler2018low,litinski2019magic}. As shown in Figure~\ref{fig:t_cost}, recent progress in magic-state cultivation~\cite{gidney2024magic} has reduced this overhead by more than 100$\times$, bringing T-gate costs close to those of logical CNOT operations. This shift also reflects an Amdahl's law effect: as the T-gate component is accelerated, the remaining Clifford component increasingly limits end-to-end improvement. Thus, once magic-state overhead becomes comparable to logical Clifford operations, further T-only optimizations yield diminishing returns unless Clifford overhead is also reduced.

For the benchmarks in Table~\ref{tab:benchmarks} (see Section~\ref{sec:benchmarks} for details), which cover key FTQC algorithms, Clifford gates now account for 58--65\% of execution overhead, with an average of 60.5\%. From an Amdahl's law perspective, even eliminating the T-attributable portion entirely would provide only a bounded speedup of roughly $1.7\times$--$2.0\times$ in this regime. As T-gate costs continue to fall, Clifford overhead, once considered negligible, has emerged as a dominant challenge to scaling FTQC.

\subsection{The Parallelism Trade-Off for Tackling Clifford Overhead}
\label{sec:pbc_source_dependency}

A common method for reducing Clifford overhead is Pauli-Based Compilation (PBC), which commutes all Clifford gates to the end of the circuit and absorbs them into the final measurements. While this eliminates explicit Clifford gates, it comes at the significant cost of reduced gate-level parallelism.

\begin{figure}[h!]
    \centering
    \includegraphics[width=\linewidth]{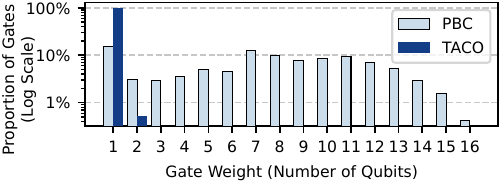}
    \caption{Gate-weight distribution of PBC and \ours for the 20-qubit QFT circuit. PBC produces multi-qubit operations involving up to 16 qubits, while \ours keeps all operations at one- or two-qubit weight. This preserves high gate parallelism by avoiding wide, multi-qubit gates.}
    \label{fig:op_weights}
\end{figure}

Commuting single-qubit Clifford gates through non-Clifford gates keeps non-Clifford operations localized to individual qubits, though in altered forms. However, when two-qubit Clifford gates are commuted through non-Clifford gates, these operations become entangled with additional qubits, resulting in multi-qubit rotation gates. This process compounds as more two-qubit Cliffords are moved, ultimately generating operations that act on an increasing number of qubits, i.e., with a higher weight. Figure~\ref{fig:op_weights} shows the gate-weight distribution from PBC for the 20-qubit QFT circuit. Starting from only single- and two-qubit gates, PBC generates high-weight operations involving up to 16 qubits.

This effect is particularly problematic for quantum algorithms, which depend on multi-qubit entanglement. After PBC transformation, most non-Clifford gates act on large subsets of qubits, severely restricting parallel execution opportunities. As shown in Figure~\ref{fig:framework_preview}(c), the 40,777 gates produced by PBC for the 18-qubit QFT circuit yield a circuit depth of 37,756, averaging just 1.08 gates per circuit layer.

\subsection{Preserving Parallelism with Structured Clifford+T Circuits}

A key challenge in reducing Clifford overhead without losing parallelism lies in how CNOT gates are treated. By deliberately avoiding the commutation of CNOT gates through the circuit, one can preserve the natural gate parallelism critical for efficient quantum computation. Thereafter, one can then focus on optimizing the remaining components, namely Toffoli gates and single-qubit gate sequences, and eliminating Clifford gates within these sequences. This helps target the primary sources of Clifford overhead while maintaining parallelism.

To optimize single-qubit gate sequences, we use the Matsumoto-Amano (MA) Normal Form~\cite{matsumoto2008representation,giles2013remarks}, a canonical structure for any single-qubit Clifford+$T$ sequence:

\begin{equation}
\text{MA Normal Form}\coloneq (T|\epsilon)\,(HT|SHT)^*\,C
\label{eq:normal_form}
\end{equation}

In this form, $(T|\epsilon)$ is an optional initial T gate, followed by a sequence of HT or SHT patterns (denoted by $(HT|SHT)^{*}$), ending with a Clifford gate $C$. Notably, the MA Normal Form guarantees both a minimal T-gate count and a unique decomposition for any target unitary, enabling systematic and efficient optimization of these sequences. By converting all single-qubit gate sequences into MA Normal Form, one can isolate Clifford reduction to two subproblems. First, eliminating redundant Cliffords within Toffoli decompositions, and second, those within these structured single-qubit sequences. This targeted strategy enables effective Clifford gate removal while preserving circuit parallelism. As shown in Figure~\ref{fig:op_weights}, \ours limits operation weight to at most two qubits.

\section{Design of \ours}
\label{sec:design}

\begin{figure}[h!]
    \centering
    \includegraphics[width=0.8\linewidth]{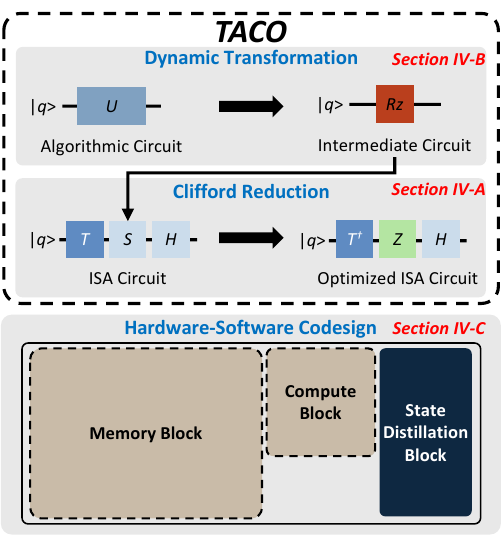}
    \caption{Overview of the design. \ours begins with dynamic gate decomposition of the algorithmic circuit, transforming it into an intermediate form with reduced \(Rz\) gates before synthesizing into a Clifford+T circuit. \ours then applies Clifford reduction to minimize execution time. The resulting circuit's high locality of \(\pi/4\) rotations informs a software-guided hardware design.}
    \label{fig:design_overview}
\end{figure}

\ours is a full-stack optimization framework for FTQC. As shown in Figure~\ref{fig:design_overview}, \ours unifies dynamic circuit decomposition, Clifford gate reduction, and architecture-aware optimization in a single workflow. We begin by presenting techniques for Clifford gate reduction, both for single-qubit sequences (Section~\ref{sec:design_clifford_reduction}) and CCX (Toffoli) gate decompositions (Section~\ref{sec:ccx_clifford_reduction}). Next, we introduce a dynamic circuit transformation pass (Section~\ref{sec:design_dynamic_transform}) that efficiently minimizes high-cost gates during synthesis, leveraging circuit structure for further reductions. Finally, we show how the locality patterns emerging from these optimized circuits directly inform the design of a tailored FTQC architecture (Section~\ref{sec:design_architecture}).

\subsection{Structured Clifford Reduction for Single-Qubit Gates}
\label{sec:design_clifford_reduction}

\ours systematically identifies all single-qubit gate sequences in the circuit and converts them into Matsumoto-Amano (MA) Normal Form (Equation~\ref{eq:normal_form}). This structured representation enables \ours to apply Clifford reduction to each sequence, effectively eliminating most Clifford gates while preserving circuit integrity. The approach is shown below with a representative gate sequence:

\begin{equation}
T\ S\ H\ T\ H\ T\ S\ H\ T 
\label{eq:sample_sequence}
\end{equation}

\subsubsection{Eliminating Phase Gates}
\label{sec:remove_s}

In MA Normal Form, only two fundamental gate patterns can appear -- $HT$ and $SHT$, which result in four possible gate patterns in the sequence:

\begin{equation}
    HT\ HT; \ HT\ SHT;\ SHT\ HT;\ SHT\ SHT
\end{equation}

\begin{figure}[b]
    \centering
    \includegraphics[width=0.85\linewidth]{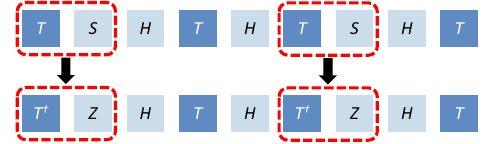}
    \caption{Remove Phase (S) gate using the identity: $TS = T^\dagger Z$.}
    \label{fig:remove_s}
\end{figure}

A key observation is that \emph{every S gate in these sequences is immediately preceded by a T gate}, with a single exception: an initial \(SHT\) prefix. An equivalence exists between TS gates and T$^\dagger$Z gates, which we can verify through:

\begin{equation}
\setlength{\arraycolsep}{2pt}
TS\!=\!
\begin{bmatrix}\mstrut 1&0\\0&e^{i\tfrac{\pi}{4}}\end{bmatrix}
\begin{bmatrix}\mstrut 1&0\\0&i\end{bmatrix}
\!=\!
\begin{bmatrix}\mstrut 1&0\\0&e^{i\tfrac{\pi}{4}}\!i\end{bmatrix}
\!=\!
\begin{bmatrix}\mstrut 1&0\\0&e^{i\tfrac{3\pi}{4}}\end{bmatrix}
\end{equation}

\begin{equation}
\setlength{\arraycolsep}{2pt}
T^\dagger Z\!=\!
\begin{bmatrix}\mstrut 1&0\\0&e^{-i\tfrac{\pi}{4}}\end{bmatrix}
\begin{bmatrix}\mstrut 1&0\\0&-1\end{bmatrix}
\!=\!
\begin{bmatrix}\mstrut 1&0\\0&-e^{-i\tfrac{\pi}{4}}\end{bmatrix}
\!=\!
\begin{bmatrix}\mstrut 1&0\\0&e^{i\tfrac{3\pi}{4}}\end{bmatrix}
\end{equation}

Note that T$^\dagger$ and Z gates are natively supported on hardware, with Z gates being executed virtually. With this equivalence, we can effectively transform all expensive S gates into `free' Z gates. The sample gate sequence shown in Equation~\ref{eq:sample_sequence} can be transformed into the new sequence shown in Figure~\ref{fig:remove_s}.

\subsubsection{Eliminating All Pauli Gates}
\label{sec:remove_pauli}
\begin{figure}[h!]
    \centering
    \includegraphics[width=0.85\linewidth]{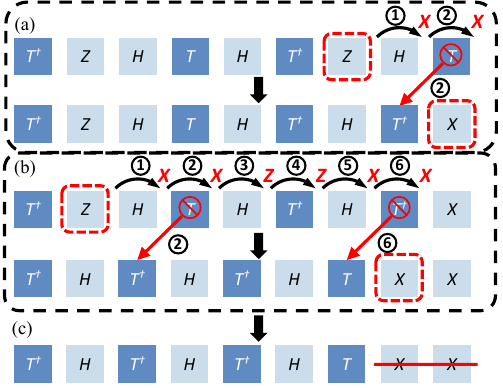}
    \caption{Commute and merge Pauli gates using commuting rules defined in Equation~\ref {eq:remove_pauli}. (a,b) shows the commuting process of the two Pauli gates and (c) the merging of the Pauli gates. Circled numbers are the step order.}
    \label{fig:remove_pauli}
\end{figure}

Once we eliminate all S gates from the sequence, we are left with H, T, T$^\dagger$, and Z gates. Although Z gates can be executed virtually, we want to eliminate them first to make the next step of eliminating H gates easier. The following commutation relations hold:

\begin{align}
    ZH = HX;\ XH = HZ \nonumber\\
    ZT = TZ;\ ZT^\dagger = T^\dagger Z \nonumber\\
    XT = T^\dagger X ;\ XT^\dagger = TX
    \label{eq:remove_pauli}
\end{align}

These relations enable all the Pauli gates to be commuted at the end of the sequence and merged as shown in Figure~\ref{fig:remove_pauli}.

\subsubsection{Eliminating All Hadamard Gates}
\label{sec:remove_h}

To this point, the remaining gates in the sequence are H, T, and T$^\dagger$ with a potential Pauli gate at the end of the sequence. Unlike Pauli Gates, Hadamard cannot be easily commuted to the end of the sequence. To further eliminate Hadamard gates, we introduce a new gate operator: \(Rx(\frac{\pi}{4})\). This operator requires the same hardware resources and implementation complexity as the T gate, with the only distinction being that the T gate performs lattice surgery between a magic state and the Z edge of the target qubit, whereas \( R_x\left(\frac{\pi}{4}\right) \) operates on the X edge~\cite{horsman2012surface,litinski2019game}. The following commutation relation holds:

\begin{equation}
   HT = Rx(\frac{\pi}{4})H,\hspace{0.1in} HT^\dagger = Rx^\dagger(\frac{\pi}{4})H,\hspace{0.1in} HH = I
    \label{eq:remove_h} 
\end{equation}

\begin{figure}[h!]
    \centering
    \includegraphics[width=0.85\linewidth]{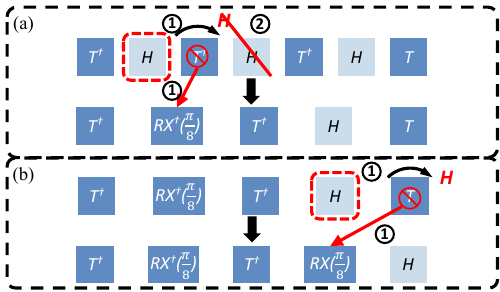}
    \caption{Commuting all the Hadamard gates to the end of the sequence using the commuting rules in Equation~\ref{eq:remove_h}. Circled numbers indicate the step order.}
    \label{fig:remove_h}
\end{figure}

We iterate over the sequence from left to right to efficiently remove all Hadamard gates. When encountering an \(H\), we commute it forward using Equation~\ref{eq:remove_h}. The two cancel if we meet a second \(H\). This process is repeated until all \(H\) gates are pushed to the end or eliminated. Figure~\ref{fig:remove_h} illustrates this.

\subsubsection{Transpilation Efficiency}
\label{sec:overhead}

The three FTQC-specific optimizations, elimination of Phase~(Section~\ref{sec:remove_s}), Pauli~(Section~\ref{sec:remove_pauli}), and Hadamard gates~(Section~\ref{sec:remove_h}), each run in linear time $O(n)$, where $n$ is the number of gates in the circuit. These passes are highly efficient in practice. For example, our 18-qubit QFT benchmark (see Section~\ref{sec:casestudy_pbc}) contains 459 $R_z$ gates and expands to 104,217 Clifford+T gates after synthesis. \ours completes the entire transpilation in under 1 second, compared to 16.1 seconds required by GridSynth for gate synthesis alone. Thus, \ours introduces negligible overhead and imposes no FTQC compilation bottleneck.

\subsubsection{Exploiting Locality in Optimized Circuits}
\label{sec:locality}

Following the three Clifford+T gate sequence optimizations, the resulting circuit displays a high degree of \textbf{locality}. That is, multiple \(R\left(\frac{\pi}{4}\right)\) rotations are often applied consecutively to the same qubit. Because each such rotation requires interaction with a magic state, this structural locality can be exploited to improve execution efficiency by strategically placing qubits near magic-state sources. This intrinsic property of the optimized circuit directly informs our architecture design in Section~\ref{sec:design_architecture}, enabling more efficient resource allocation and throughput.

\subsection{Parallelism-Preserving Clifford Reduction in Toffoli Gates}
\label{sec:ccx_clifford_reduction}
\begin{figure}[h!]
   \centering
   \includegraphics[width=\linewidth]{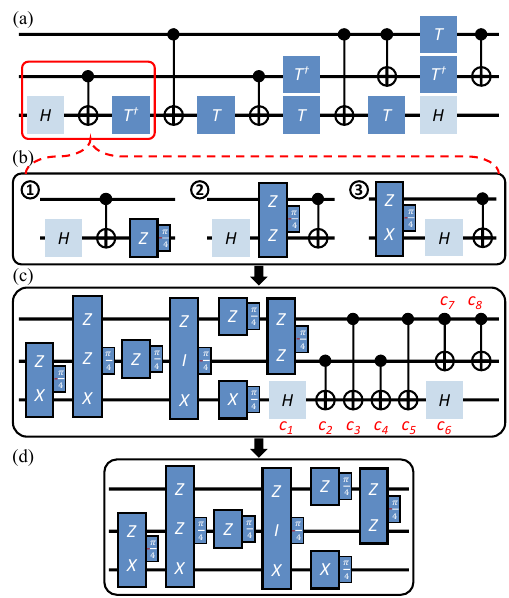}
   \caption{(a) Decomposition of the Toffoli gate into Clifford and non-Clifford components. (b) Commutation of the first non-Clifford gate ($T^{\dagger}$): \textcircled{1} $T^{\dagger}$ represented as $R_Z(-\pi/4)$, \textcircled{2} commuted through a CNOT, and \textcircled{3} through a Hadamard gate. (c) The circuit obtained after commuting all non-Clifford gates to the front. (d) Eight Clifford gates cancel pairwise after swapping $C_3$ and $C_4$. Consecutive self-inverse Clifford pairs ($C_2$, $C_4$), ($C_3$, $C_5$), and ($C_7$, $C_8$) cancel first, followed by the remaining pair ($C_1$, $C_6$).}
   \label{fig:ccx_optimization}
\end{figure}

A single Toffoli gate decomposes into seven $T/T^{\dagger}$ gates, six CNOTs, and two Hadamard gates, as shown in Figure~\ref{fig:ccx_optimization}(a). Figure~\ref{fig:ccx_optimization}(b) illustrates the commutation process of the first non-Clifford gate ($T^{\dagger}$). \textcircled{1}: the $T^{\dagger}$ gate is represented as $R_Z(-\pi/4)$. \textcircled{2}: it is then commuted through a CNOT gate, and \textcircled{3}: subsequently through a Hadamard gate.

Figure~\ref{fig:ccx_optimization}(c) shows the circuit after all remaining non-Clifford gates have been commuted to the front, leaving the Clifford gates grouped afterward and labeled $C_1$ through $C_8$. These Clifford gates can be locally canceled. Gates $C_3$ and $C_4$, both CNOTs acting on the same target qubit, commute and can be swapped. After this swap, the consecutive self-inverse pairs $(C_2, C_4)$, $(C_3, C_5)$, and $(C_7, C_8)$ cancel, followed by the remaining Hadamard pair $(C_1, C_6)$.

After these local cancellations, only the non-Clifford rotations acting on the three qubits of the original Toffoli gate remain, as shown in Figure~\ref{fig:ccx_optimization}(d). All operations stay confined to this local three-qubit subspace, ensuring that no additional commutation steps or circuit depth overhead are introduced.

\subsection{FTQC-oriented Dynamic Circuit Transformation}
\label{sec:design_dynamic_transform}

\begin{figure}[b]
   \centering
   \includegraphics[width=0.9\linewidth]{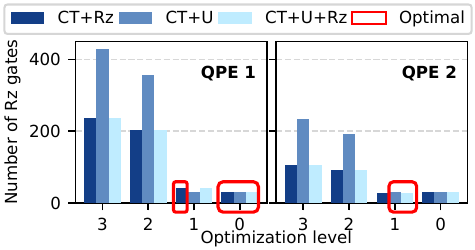}
   \caption{Number of Rz gates in intermediate circuits obtained from transpiling two 4-qubit QPE circuits using Qiskit Transpiler across optimization levels 0-3 and three basis gate sets (Clifford+T+U, Clifford+T+Rz, Clifford+T+U+Rz). Lower Rz gate counts indicate better optimization.}
   \label{fig:qpe_qiskit_rzs}
\end{figure}

A key step in our workflow is an intermediate circuit transformation, motivated by the fact that efficient gate synthesis algorithms such as GridSynth~\cite{ross2014optimal} operate on a restricted gate set. Reducing the number of gates that require synthesis at this stage directly impacts overall resource cost. To highlight this challenge, we evaluated common NISQ-oriented transpilers on two 4-qubit Quantum Phase Estimation (QPE) circuits. As shown in Figure~\ref{fig:qpe_qiskit_rzs}, no configuration consistently achieves low $R_z$ gate counts for practical FTQC needs, revealing an important gap in current transpiler strategies.

To address this, we develop a dynamic transformation method optimized for FTQC that consists of three stages. First, we study all possible decomposition rules for gates up to three qubits, selecting the option with the minimum FTQC cost for each. Second, we simplify trivial cases by replacing rotation gates that are Clifford+T equivalent, such as mapping $R_z(\pi)$ to a Pauli-Z. Third, we merge consecutive single-qubit gates whenever possible, leveraging local cancellations to further reduce the gate count. This targeted approach yields optimal gate counts on both QPE benchmarks, consistently outperforming NISQ-oriented transpiler configurations and demonstrating the value of FTQC-specific dynamic transformation.

\subsection{Architecture Co-Design Guided by Circuit Locality}
\label{sec:design_architecture}

The optimized Clifford+T circuits from Section~\ref{sec:design_clifford_reduction} exhibit high \(\pi/4\) rotation locality -- that is, long sequences of \(R\left(\frac{\pi}{4}\right)\) rotations are repeatedly applied to the same qubit. The length of these sequences can span dozens or even hundreds of consecutive rotations. To exploit this structural regularity, we propose an FTQC architecture consisting of two logical regions: a \textit{compute block} and a \textit{memory block}. The compute block hosts logical qubits that undergo frequent $\pi/4$ rotations, enabling efficient interaction with magic-state resources. The memory block holds the remaining qubits in a compact, area-efficient layout while still supporting necessary Clifford operations such as CNOT and Hadamard gates.

\subsubsection{Compute Block}
\label{sec:design_compute_block}

The compute block is optimized for executing long sequences of $\frac{\pi}{4}$-rotation operations. In the Clifford-Reduced circuit from Section~\ref{sec:design_clifford_reduction}, there are four types of $\frac{\pi}{4}$ rotations: Z-axis rotations, X-axis rotations, and their inverses. The $Rz(\frac{\pi}{4})$ gate, equivalent to the T gate, requires lattice surgery between a magic state qubit and the Z-edge of the target qubit. Similarly, the $Rx(\frac{\pi}{4})$ gate requires interaction with the X-edge of the target qubit.

\begin{figure}[h!]
    \centering
    \includegraphics[width=0.9\linewidth]{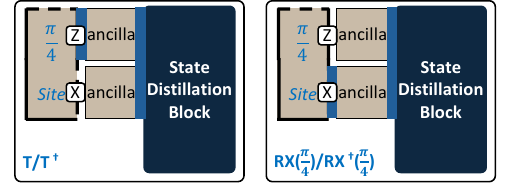}
    \caption{Layout of the compute block, showing the $\frac{\pi}{4}$ logical qubit site with both X and Z edges exposed via ancilla qubits to the state distillation block. This configuration supports $Rz(\frac{\pi}{4})$ (left) and $Rx(\frac{\pi}{4})$ rotations (right) by enabling selective edge interaction. One compute block uses four logical qubit tiles: two for the $\frac{\pi}{4}$-site qubit and two ancilla tiles.}
    \label{fig:compute_block}
\end{figure}

As shown in Figure~\ref{fig:compute_block}, each compute block includes a \(\pi/4\) logical qubit site with X and Z edges exposed through ancilla tiles connected to the state distillation block. These edges are activated according to the required rotation axis. During magic-state injection, the sign of the resulting \(R(\frac{\pi}{4})\) rotation is probabilistic. If the opposite sign is produced, the intended rotation is recovered using a Clifford correction, which is unavoidable but much cheaper than another non-Clifford rotation. High \(\pi/4\)-rotation locality ensures that each target qubit remains in the compute block throughout its rotation sequence, avoiding unnecessary movement between memory and compute regions and improving execution efficiency.

\subsubsection{Memory Block}
\label{sec:design_mem_block}
\begin{figure}[h!]
    \centering
    \includegraphics[width=0.9\linewidth]{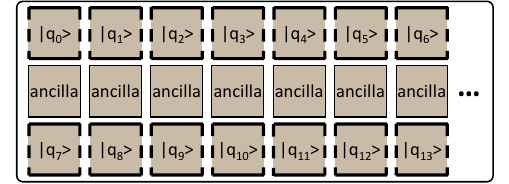}
    \caption{Memory block layout for storing idle qubits. Logical qubits occupy the first and third rows, while the middle row is reserved for ancilla qubits.}
    \label{fig:memory_block}
\end{figure}

The memory block stores the remaining logical qubits, which are mostly idle except during the execution of CNOT or Hadamard gates. We adopt the compact tile layout from~\cite{litinski2019game}, illustrated in Figure~\ref{fig:memory_block}. The design consists of three rows: logical qubit tiles occupy the first and third rows, while the middle row contains ancilla tiles. This arrangement requires only \(1.5n\) tiles for \(n\) logical qubits and supports direct interactions between any qubit pair via the shared ancilla layer, enabling flexible and efficient CNOT operations. Hadamard gates can be applied locally with minimal overhead. The central ancilla row is also connected to the compute block, enabling seamless movement of qubits between memory and compute regions as needed for fault-tolerant execution.

\subsubsection{Patch Rotation}
In the memory block, each logical qubit exposes only a single computational basis for interactions. When operations require the complementary basis, a \textit{patch rotation} is performed: 1 cycle to expand the patch, one cycle to rotate the logical basis, and one cycle to shrink back, totaling 3 code cycles~\cite{litinski2019game}. Since these rotations are infrequent, they do not impact overall performance.

\subsubsection{Qubit Transfer Between Memory and Compute Blocks}
When a logical qubit moves between memory and compute blocks, the architecture provides direct access from any memory location to the compute block. The transfer completes in a single code cycle by expanding the qubit into the compute block and contracting it back to its operational configuration.

\subsection{Optimizing Compute and Distillation Block Placement}
\label{sec:multi_blocks}
\begin{figure}[h]
    \centering
    \includegraphics[width=\linewidth]{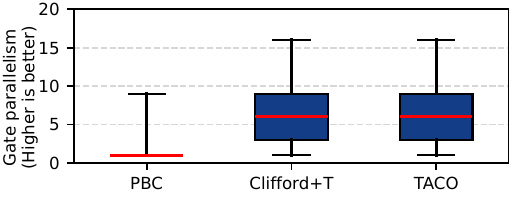}
    \caption{Gate parallelism comparison of 20-qubit QFT. TACO preserves the same high gate-level parallelism as the original Clifford+T circuit.}
    \label{fig:gate_parallelism}
\end{figure}

As shown in Figure~\ref{fig:gate_parallelism}, \ours-optimized circuit preserves the high gate-level parallelism of the original Clifford+T circuit. This parallelism requires multiple magic states within each circuit layer, which necessitates multiple distillation blocks. This raises two practical questions: how many distillation blocks are optimal, and where should they be placed.

\begin{figure}[h!]
    \centering
    \includegraphics[width=\linewidth]{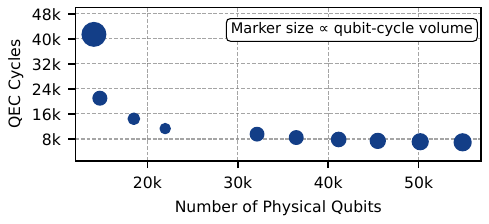}
    \caption{Qubit-cycle comparison for 18-qubit QFT execution with magic state preparation throughputs from 1-10 magic states/cycle using Magic State Cultivation~\cite{gidney2024magic}. Marker sizes are proportional to qubit-cycle volume, with four magic states/cycle being the optimal configuration.}
    \label{fig:qubit_cycle_tradeoff}
\end{figure}

\noindent \textbf{Minimizing Space-Time Volume via Factory Scaling.}
The primary architectural objective of \ours is to minimize the total space-time volume. This requires balancing the physical qubit overhead of magic state factories against the resulting increase in circuit throughput. Unlike prior heuristics that suggest fixed overhead ratios for distillation, we explicitly model the physical cost of each distillation block based on its surface code distance and layout requirements. Figure~\ref{fig:qubit_cycle_tradeoff} illustrates the QEC cycles required for an 18-qubit QFT circuit as a function of available magic state blocks. The marker size indicates the total qubit-cycle volume, which incorporates the actual spatial footprint of both memory qubits and the allocated distillation factories. We observe that as throughput increases, the cycle count decreases significantly, reaching an optimal volume at 4 magic-state blocks. This configuration represents a 57\% reduction in volume compared to a single-block setup. Beyond this point, further scaling yields diminishing returns because the added spatial overhead outweighs the marginal cycle reductions. \ours automatically identifies this optimal configuration by simulating these trade-offs, ensuring that resource allocation is driven by total volume minimization rather than arbitrary overhead limits.

\begin{figure}[h!]
    \centering
    \includegraphics[width=\linewidth]{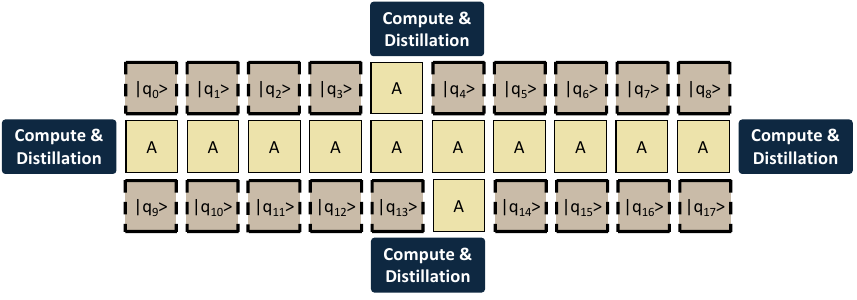}
    \caption{Architecture for 18-qubit QFT execution featuring 18 logical memory qubits and 4 distributed compute and distillation blocks.}
    \label{fig:qft18_arhcitecture}
\end{figure}

\noindent \textbf{Spatial Organization of Compute and Distillation Blocks.}
To achieve a target throughput of one magic state per cycle per compute block, we group compute blocks with multiple distillation units to form a ``super-block.'' This organization is motivated by the fact that individual distillation protocols often exhibit lower throughput than the consumption rate of a compute block~\cite{litinski2019magic}. These super-blocks are distributed around the central memory qubits to minimize movement overhead and routing congestion, while ensuring every compute block can access any logical qubit via shared ancilla paths. Figure~\ref{fig:qft18_arhcitecture} demonstrates this layout for the 18-qubit QFT. 

This strategy is informed by the broader principle that modern architectural optimizations have significantly reduced the resource cost per magic state~\cite{litinski2019magic}. By treating the factory cost as a primary variable in our volume-minimization strategy rather than a static overhead, \ours can adaptively scale distillation resources. This enables higher throughput and shorter execution times while maintaining a favorable balance between hardware investment and operational speed.

\subsection{Physical Realization of the Proposed Architecture}
\label{sec:specialized_layout}

The proposed architecture can be physically realized as a specialized lattice-surgery organization built on standard surface-code patches, similar to prior compact, intermediate, and fast FTQC organizations~\cite{litinski2019game}. The key difference is that our organization is tailored to the high locality exposed by \ours-optimized circuits.

We distinguish between \emph{compute} regions and \emph{memory} regions. This distinction is architectural rather than technological: both regions use the same surface-code substrate, physical qubits, and couplers, but serve different logical roles during execution. Compute regions handle active non-Clifford processing, while memory regions primarily store logical states and support Clifford operations. To support arbitrary $\pi/4$ rotations, compute regions employ a modified logical patch that has the same code distance $d$ while expanding the physical footprint from the standard $d \times d$ patch to approximately $2d \times d$, requiring roughly $2\times$ more physical qubits per logical patch. In contrast, memory regions retain the standard patch structure.

Under this organization, qubits are moved between memory and compute regions by teleporting or swapping the logical state between physical patch locations, while the underlying physical qubits remain unchanged. This can be implemented using standard lattice-surgery primitives.

\section{Evaluation Methodology}

\subsection{Figure of Merit}
The primary metric used to evaluate \ours is the reduction in Clifford gate counts. We specifically measure the percentage decrease in \textsc{cnot}, Hadamard ($H$), and Phase ($S$) gates. Additionally, we estimate potential execution time savings by calculating QEC cycle reductions based on the logical gate latencies detailed in Section~\ref{sec:background_spact_time_cost}.

\subsection{QEC Cycle Calculation}
\label{sec:methodology_qec_cycles}

We evaluate the temporal overhead of FTQC circuits using two complementary methodologies to capture both theoretical lower bounds and practical architectural constraints.

\vspace{0.5em}
\noindent \textbf{Architecture-Independent Analysis.} 
We first calculate the total QEC cycles as the critical path length, utilizing the cycles-per-gate values from Section~\ref{sec:background_spact_time_cost}. This metric represents the theoretical minimum cycle count based on the longest dependency chain, independent of hardware resource limitations or routing overhead.

\vspace{0.5em}
\noindent \textbf{Cycle-Accurate Simulation and Routing.} 
To capture realistic architectural constraints, we use a cycle-accurate simulator that processes the circuit layer by layer. While local operations are applied directly, non-local gates (\textsc{cnot}) and state-injections ($T$, $S$) are routed to available compute blocks. By default, our simulator employs \textbf{greedy routing}, which allocates resources to ready gates on a first-come, first-served basis. As a specialized case study demonstrating compatibility with advanced resource management, we integrate \textbf{LSQECC routing}~\cite{watkins2024high}. This dual-routing approach validates \ours's performance across varying architectural sophistication.

\subsection{Benchmark Circuits}
\label{sec:benchmarks}

We evaluate \ours using a diverse set of benchmarks categorized by their structural characteristics. First, we select key algorithms from QASMBench~\cite{li2022qasmbenchlowlevelqasmbenchmark}, including the Quantum Fourier Transform (\texttt{qft}), Quantum Phase Estimation (\texttt{qpe}), \texttt{ising} model simulation, and \texttt{w\_state} preparation. 

Second, we include four Toffoli-heavy FTQC algorithms from Op-T-mize~\cite{Kottmann2024_op-T-mize}: \texttt{adder}, \texttt{csla\_mux}, \texttt{hwb}, and \texttt{qcla\_mod}. These represent essential arithmetic subroutines for high-level algorithms such as Shor's algorithm~\cite{shor1999polynomial}. Table~\ref{tab:benchmarks} summarizes the qubit and gate counts for these benchmarks.

\begin{table}[h]
\centering
\caption{Benchmark Circuit Characteristics.}
\label{tab:benchmarks}
\resizebox{\columnwidth}{!}{%
\begin{tabular}{c|cccccccc}
\hline
& \textbf{qft}  & \textbf{ising}  & \textbf{qpe} & \textbf{w\_state} & \textbf{adder} & \textbf{csla\_mux}  & \textbf{hwb} &  \textbf{qcla\_mod}  \\
\hline
\textbf{Qubits} & 18  & 26  & 9 & 76 & 24 & 15 & 16 & 26 \\
\textbf{Gates} & 783 & 307  & 36 & 378 & 330 & 70 & 31764 & 294 \\
\hline
\textbf{Clifford+$T$} 
& \multirow{2}{*}{95968} 
& \multirow{2}{*}{25520}   
& \multirow{2}{*}{7587} 
& \multirow{2}{*}{38223} 
& \multirow{2}{*}{1128} 
& \multirow{2}{*}{210} 
& \multirow{2}{*}{91642} 
& \texttt{1120} \\
\textbf{Gates} & & & & & & & & \\
\hline
\textbf{Increase ($\times$)} & 122.6 & 83.1 & 210.8 & 101.1 & 3.4 & 3.0 & 2.9 & 3.8 \\
\hline
\end{tabular}%
}
\end{table}

\vspace{0.5em}
\noindent \textbf{Scalability and Compatibility.} 
To evaluate scalability, we utilize large-scale \texttt{qft} circuits ranging from 100 to 300 qubits. Furthermore, to demonstrate that \ours is complementary to existing FTQC synthesis frameworks, we conduct a case study on circuits synthesized by \texttt{Synthetiq}~\cite{paradis2024synthetiq} and \texttt{TRASYN}~\cite{hao2025reducing}. We apply \ours as a post-processing step to these outputs to further optimize final gate counts.

\vspace{0.5em}
\noindent \textbf{Expansion to Clifford+$T$ Basis.} 
Table~\ref{tab:benchmarks} reports both the original gate counts (\textbf{Gates}) in native representation and compiled counts in the Clifford+$T$ basis. Notably, rotation-heavy circuits (e.g., \texttt{qft}, \texttt{qpe}) exhibit an expansion of over $83\times$, while Toffoli-dominated circuits experience a smaller growth of $3\text{--}4\times$. This disparity arises from the decomposition of high-level operations: a single rotation gate is synthesized into hundreds of Clifford+$T$ gates, of which roughly $60\%$ are Clifford and $40\%$ are $T$ gates. In contrast, a Toffoli gate can be decomposed into 8 Clifford gates and 7 $T$ gates. Consequently, in rotation-dominated circuits, over $99\%$ of the Clifford+$T$ gates originate from rotation synthesis, whereas for Toffoli-heavy circuits, $\sim 70\%$ result from Toffoli decomposition.

\subsection{Clifford+$T$ Synthesis}
\label{sec:circuit_synthesis}

We compare \ours against the native Clifford+$T$ transpilation functionality in Qiskit~2.2.3 (Optimization Level 3). While Qiskit employs the Solovay-Kitaev algorithm~\cite{dawson2005solovay} for approximate unitary synthesis, \ours leverages the GridSynth algorithm~\cite{ross2014optimal} for optimal $T$-count synthesis. For all cases, we set a default synthesis error tolerance of $\epsilon = 10^{-10}$.

\section{Evaluation Results}

\subsection{Clifford Gate Reduction}
\label{sec:clifford_reduction}

\begin{figure}[h!]
    \centering
    \includegraphics[width=\linewidth]{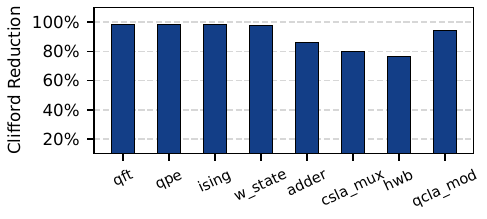}
    \caption{Clifford gate reduction percentage across benchmark circuits. \ours achieves an average 91.2\% Clifford gate reduction across all benchmarks.}
    \label{fig:clifford_reductions}
\end{figure}

Figure~\ref{fig:clifford_reductions} presents Clifford gate reduction results for all benchmark circuits. \ours achieves an average Clifford gate reduction of 91.2\% and up to 98.6\%, highlighting the strength of our optimization. Notably, QFT and QPE circuits show reductions of 98.6\% and 98.1\%, respectively. The smallest reduction is observed in the hwb circuit at 77\%, due to its high initial CNOT gate content (over 65\%). Since \ours preserves CNOT gates to maintain gate parallelism, the potential for Clifford reduction in hwb is inherently limited. Nevertheless, as shown in Section~\ref{sec:speedups}, retaining these CNOT gates ultimately improves overall execution runtime.

\subsection{Speedup of \ours over PBC}
\label{sec:speedups}

\begin{figure}[h!]
    \centering
    \includegraphics[width=\linewidth]{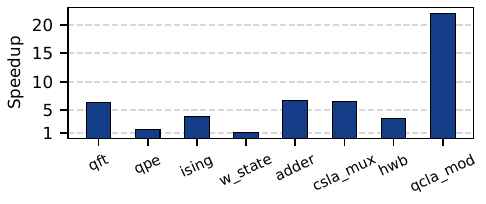}
    \caption{\ours achieves 1.14$\times$ to 21.9$\times$ speedup across benchmark circuits over PBC, with a geometric mean speedup of 4.4$\times$.}
    \label{fig:taco_speedup}
\end{figure}

With enhanced gate parallelism, \ours achieves significant speedup over PBC across the benchmarks as shown in Figure~\ref{fig:taco_speedup}. On average, \ours achieves 4.4$\times$ speedup over the baseline approach. The smallest speedup is observed for w\_state at 1.14$\times$, which can be attributed to the circuit's limited CNOT gate density of only 1\%. As discussed in Section~\ref{sec:pbc_source_dependency}, the primary source of limited parallelism stems from commuting CNOT gates through the circuit. Since w\_state contains relatively few CNOT gates, its circuit depth remains comparable between PBC and \ours, resulting in similar QEC runtime with only modest improvement. Interestingly, the hwb circuit, which has a high CNOT ratio and shows the worst Clifford gate reduction results, still achieves a 3.52$\times$ speedup, demonstrating that runtime performance can be significantly improved even when gate-reduction opportunities are limited.  For all remaining benchmark circuits, \ours achieves at least 1.54$\times$ speedup, with a maximum of 21.9$\times$ for circuits with higher CNOT gate density and greater opportunities for parallel execution. 

\subsection{\ours{} versus PBC: Overall FTQC Volume Reduction}
\label{sec:casestudy_pbc}

In this section, we perform a comprehensive FTQC resource estimation comparison between PBC and \ours using the 18-qubit QFT circuit. Both approaches produce optimized circuits with 40,777 T gates. While PBC eliminates all Clifford gates, \ours retains 1,080 Clifford gates, resulting in total gate counts of 40,777 and 41,857 gates, respectively. Despite this slightly higher gate count, the key advantage of \ours lies in execution parallelism: it achieves a circuit depth of 6,598 compared to PBC's 39,055, making \ours $5.9\times$ shallower and dramatically reducing the total execution cycles required.

\subsubsection{Resource Estimation Methodology} We evaluate resource requirements using established FTQC estimation methods~\cite{litinski2019game, gidney2021factor, gidney2024magic}. We assume a physical error rate of $10^{-3}$ and compare two PBC architectures: a compact, area-optimized design and a fast design. We select distillation protocols targeting T-gate error below 1\%, with code distance chosen to maintain logical error below 1\%. These parameters are standard in FTQC resource estimations~\cite{litinski2019game, gidney2021factor}.

\begin{figure}[h!]
    \centering
    \includegraphics[width=\linewidth]{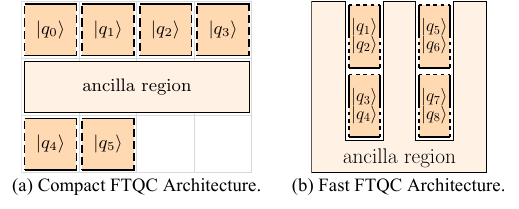}
    \caption{PBC architectures for comparison.}
    \label{fig:ftqc_arch}
\end{figure}

\subsubsection{PBC Architecture} We compare \ours against PBC using two FTQC architectures from prior work~\cite{litinski2019game}. In PBC, multi-qubit $\frac{\pi}{4}$ rotations are executed via lattice surgery between logical qubits and magic state ancillae. Each $\frac{\pi}{4}$ rotation acts on logical edges set by the Pauli string. For example, a \texttt{ZZZY} Pauli string requires lattice surgery on the Z edges of qubits 0-2 and both X and Z edges of qubit 3.

Figure~\ref{fig:ftqc_arch} shows the two baseline architectures. The `Compact design' uses $1.5n + 3$ qubit tiles but exposes only one edge per qubit to the ancilla region. When the required edges are not exposed, qubits must be rotated before surgery can proceed. Y-edge operations are particularly costly, requiring simultaneous exposure of both X and Z edges. This necessitates three additional ancilla qubits and up to 9 cycles per multi-qubit $\frac{\pi}{4}$ rotation. The `Fast architecture' exposes all edges of all qubits, enabling one multi-qubit $\frac{\pi}{4}$ rotation per cycle at the cost of significantly more tiles ($2n + \sqrt{8n}+1$).

\subsubsection{Distillation Protocol and Throughput} With $4 \times 10^4$ T gates, each magic state requires an error rate below $0.01 / (4 \times 10^4) = 2.5 \times 10^{-7}$. We employ the standard 15-to-1 distillation protocol, which suppresses errors by $35p^3$ using 11 logical qubit tiles and produces one magic state every 11 cycles~\cite{litinski2019game}. At a physical error rate of $10^{-3}$, this yields magic states with error rates of $35 \times (10^{-3})^3 = 3.5 \times 10^{-8}$.

The required distillation blocks depend on each architecture's magic state consumption rate. PBC Compact consumes one magic state every nine cycles, requiring only one distillation block, totaling 28,783,535 QEC cycles. PBC Fast consumes one magic state per cycle, requiring 11 distillation blocks to meet demand and totaling 2,382,355 cycles. With magic state distillation blocks, the optimal number of blocks is found to be 3, each containing 11 distillation units. Using the cycle simulator (discussed in Section~\ref{sec:methodology_qec_cycles}), \ours completes execution in 760,901 cycles. This results in total logical qubit tile counts of: PBC Compact (30 data + 11 distillation = 41), PBC Fast (49 data + 121 distillation = 170), and \ours (29 data + 12 compute + 363 distillation = 404).

\begin{table}[t]
\centering
\caption{Resource comparison for 20-qubit QFT circuit}
\label{tab:resource_comparison}
\begin{tabular}{l@{\hskip 2.5pt}c@{\hskip 2.5pt}|c@{\hskip 2.5pt}|c}
\hline
\textbf{Metric} & \textbf{PBC Compact} & \textbf{PBC Fast} & \textbf{\ours} \\
\hline
Magic state blocks & 1 & 11 & 33 \\
Magic state tiles & 11 & 121 & 363 \\
Data tiles & 30 & 49 & 41 \\
\textbf{Total tiles} & 41 & 170 & 404 \\ \hline
\textbf{QEC cycles} & 28,783,535 & 2,382,355 & 760,901 \\ \hline 		
Code distance & 21 & 19 & 19 \\
\textbf{Physical qubits/tile} & 882 & 722 & 722 \\
\hline
Volume (data) & $7.6 \cdot 10^{11}$ & $8.4 \cdot 10^{10}$ & $2.3 \cdot 10^{10}$ \\
Volume (MSD) & $2.8 \cdot 10^{11}$ & $2.1 \cdot 10^{11}$ & $2.0 \cdot 10^{11}$ \\
Total volume & $1.0 \cdot 10^{12}$ & $2.9 \cdot 10^{11}$ & $2.2 \cdot 10^{11}$ \\
\textbf{Reduction with \ours} & \textbf{-79\%}  & \textbf{-24\%} & - \\
\hline
\textbf{QEC cycles w. MSC} & 28,783,535 & 2,382,355 & 595,604 \\ 
Volume (MSC) & $2.8 \cdot 10^{10}$ & $2.1 \cdot 10^{10}$ & $2.1 \cdot 10^{10}$ \\
Total volume w. MSC & $7.9 \cdot 10^{11}$ & $1.1 \cdot 10^{11}$ & $3.8 \cdot 10^{10}$ \\
\textbf{Reduction with \ours} & \textbf{-95\%}  & \textbf{-63\%} & - \\
\hline
\end{tabular}
\end{table}

\subsubsection{Code Distance} The logical error rate per logical qubit per code cycle can be approximated as $p_L = 0.1(100p)^{(d+1)/2}$~\cite{fowler2018low}, where $p$ is the physical error rate and $d$ is the code distance. To maintain overall logical error below 1\%, the code distance must satisfy: $\text{total\_tiles} \times \text{total\_cycles} \times d \times p_L < 0.01$. Given the tile and cycle counts, PBC Compact requires a minimum code distance of 21, while PBC Fast and \ours require only 19 due to shorter execution time. The physical qubits per logical qubit is calculated as $2*d^2$.

\subsubsection{Resource Comparison} Table~\ref{tab:resource_comparison} summarizes the complete resource breakdown across all three architectures. The results demonstrate that \ours significantly reduces qubit-cycle volume and achieves \textbf{79\%} and \textbf{24\%} reductions over PBC Compact and Fast, respectively. 

Moreover, when using more efficient magic state cultivation (MSC), \ours's optimal architecture containing 4 compute \& distillation blocks reduces QEC cycles to 595,604. It requires the same distance-19 code, with the volume for magic states reduced by an order of magnitude~\cite{gidney2024magic} compared to magic-state distillation. The result is \ours reduces the qubit-cycle volume by more than \textbf{95\%} and \textbf{63\%} compared to PBC Compact and PBC Fast, respectively.

It's worth noting that, although \ours uses more logical qubit tiles, its overall overhead is significantly lower than that of PBC-based approaches. More importantly, between the two PBC-based approaches, the fast design is more efficient than the compact design. PBC Compact's data volume overhead is 2.7$\times$ higher than magic state volume, while PBC Fast has comparable overhead. By adopting \ours, the magic-state overhead remains nearly constant while the data volume is significantly reduced, leading to an overall reduction. In this context, further reducing T gate overhead in PBC-based architectures has diminishing returns, whereas in TACO, such optimizations can be more impactful. Thus, TACO improves the effectiveness of existing T gate optimizations.

\subsubsection{Sensitivity to Magic-State Throughput}
\label{sec:throughput_sensitivity}

We further examine the same case study under different magic-state throughputs. In the default TACO configuration, the system provisions enough factories to supply 4 magic states per round, resulting in 64.5k physical qubits, 595k QEC cycles, and a 2.7$\times$ reduction in FTQC execution overhead measured as qubit-cycle volume relative to PBC Fast, which uses 44k physical qubits. Reducing the throughput to 3 magic states per round lowers the physical-qubit count to 47.6k, which is only slightly above PBC Fast, while increasing runtime to 760k QEC cycles. Reducing further to 2 magic states per round lowers the physical-qubit count to 30.8k, which is below PBC Fast, but increases runtime to 1.14M QEC cycles. Despite this area--time tradeoff, TACO still achieves 2.9$\times$ and 2.99$\times$ lower qubit-cycle volume than PBC Fast, respectively, as shown in Table~\ref{tab:throughput_sensitivity}. This shows that the benefit of TACO is preserved even under tighter qubit budgets.

\begin{table}[h]
\centering
\footnotesize
\setlength{\tabcolsep}{3pt}
\caption{TACO under reduced magic-state throughput.}
\label{tab:throughput_sensitivity}
\begin{tabular}{lccc}
\toprule
\multirow{2}{*}{\textbf{TACO config.}} &
\multirow{2}{*}{\textbf{Phys. qubits}} &
\multirow{2}{*}{\textbf{QEC cycles}} &
\textbf{Vol. red.} \\
& & & \textbf{vs.\ PBC Fast} \\
\midrule
4 magic states / round & 64.5k & 595k  & 2.70$\times$ \\
3 magic states / round & 47.6k & 760k  & 2.90$\times$ \\
2 magic states / round & 30.8k & 1.14M & 2.99$\times$ \\
\bottomrule
\end{tabular}
\end{table}

\subsection{\ours Combined with LSQECC and EDPC}
\label{sec:combined_lsqecc_edpc}

To demonstrate that the Clifford-reduced circuits produced by \ours can benefit from orthogonal routing optimizations, we compile the 20-qubit QFT circuit optimized by \ours using LSQECC~\cite{watkins2024high}. Since LSQECC currently does not support the \ours layout with compute sites, we use its default EDPC layout~\cite{leblond2024realistic}. Following the same methodology described in Section~\ref{sec:casestudy_pbc}, we obtain a final FTQC volume of \textbf{$1.33 \times 10^{11}$} with magic state distillation, corresponding to a further $1.7\times$ reduction compared to the naive routing baseline (Table~\ref{tab:lsqecc_results}). We expect further reductions in total FTQC volume once LSQECC supports \ours layout with compute sites as the high locality of the circuit can significantly reduce the number of operations required for routing.

\begin{table}[h]
\centering
\caption{Improved FTQC volume of \ours with LSQECC routing.}

\begin{tabular}{lcc}
\toprule
\textbf{Routing Method} & \textbf{FTQC Volume} & \textbf{Reduction} \\
\midrule
Naive (greedy) & $2.2 \times 10^{11}$ & Baseline \\
LSQECC (EDPC) & $1.33 \times 10^{11}$ & $1.7\times$ \\
\bottomrule
\end{tabular}
\label{tab:lsqecc_results}
\end{table}

These results confirm that \ours and routing optimizations target complementary layers of the compilation stack. While routing algorithms improve spatial scheduling, \ours focuses on circuit-level simplification that enables compatibility with such optimizations. In contrast, PBC offers no routing opportunities since each layer contains only a single operation. 

\begin{table}[h]
\centering
\footnotesize
\caption{LSQECC execution time for QFT.}
\setlength{\tabcolsep}{6pt}
\begin{tabular}{c|cc}
\hline
 & \textbf{LSQECC} & \textbf{LSQECC + \ours} \\
\hline
Lattice-Surgery Slices & 123,139 & 62,503 \\
\hline
\end{tabular}
\label{tab:lsqecc_qft}
\end{table}

On the other hand, \ours not only benefits from LSQECC's routing algorithm, but also improves LSQECC execution through its Clifford reduction. To quantify this effect, we run LSQECC on both the original and the \ours-optimized QFT circuits, using the same layout and magic-state factory configuration (i.e., the same physical-qubit budget). As shown in Table~\ref{tab:lsqecc_qft}, the original QFT circuit requires 123,139 lattice-surgery slices, whereas the \ours-optimized circuit requires only 62,503 slices. This corresponds to a nearly $2\times$ reduction in execution time. These results demonstrate that \ours and LSQECC are synergistic: LSQECC improves the routing of \ours-optimized circuits, while \ours reduces the execution time of LSQECC by simplifying the underlying circuit.

\subsection{\ours: Scalability}
\label{sec:scability}

\begin{figure}[h!]
    \centering
    \includegraphics[width=\linewidth]{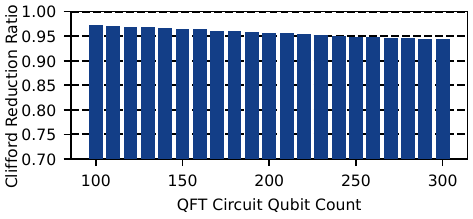}
    \caption{Clifford gate reduction ratios achieved by \ours on QFT circuits ranging from 100 to 300 qubits. The reduction ratio remains consistently around 95\% across all circuit sizes, demonstrating the scalability of \ours to large-scale fault-tolerant quantum circuits.}
    \label{fig:clifford_reduction_scaling}
\end{figure}

\ours reduces Clifford gates via local optimizations, specifically by simplifying Clifford operations in synthesized gate sequences of single-qubit $R_z$ rotations and in the decomposition of $\text{CCX}$ gates. Since these two sources dominate the Clifford gate count in FTQC circuits, the reduction achieved by \ours naturally extends to larger circuits.

To evaluate scalability, we apply \ours to QFT circuits ranging from 100 to 300 qubits and measure the resulting Clifford reduction ratios. As shown in Figure~\ref{fig:clifford_reduction_scaling}, the reduction ratio remains consistently around \textbf{95\%} across all tested sizes, demonstrating the scalability of \ours to large circuits while maintaining high optimization efficiency.

\subsection{Clifford+T Transpilation}

\begin{table}[h]
\centering
\scriptsize
\caption{Clifford+T transpilation comparison between Qiskit and \ours}
\begin{tabular}{l@{\hspace{0.2cm}}c@{\hspace{0.2cm}}c@{\hspace{0.2cm}}c|@{\hspace{0.2cm}}c@{\hspace{0.2cm}}c@{\hspace{0.2cm}}c}
\hline
& \multicolumn{3}{c}{\textbf{Qiskit-O3}} & \multicolumn{3}{c}{\textbf{\ours}} \\
\cline{2-4} \cline{5-7}
\textbf{Circuit} & \textbf{Unitaries to} & \textbf{T Gates} & \textbf{Time (s)} & \textbf{Unitaries to} & \textbf{T Gates} & \textbf{Time (s)} \\
& \textbf{Synthesize} & & & \textbf{Synthesize} & & \\
\hline
QFT       & 408 & 2,623,881 & 34.73 & 378 & 9,529  & 0.041 \\
QPE       & 30  & 275,356   & 3.39  & 30  & 411    & 0.013 \\
Ising     & 75  & 589,669   & 7.69  & 75  & 1,500  & 0.049 \\
W-State   & 148 & 1,383,832 & 18.85 & 75  & 2,218  & 0.121 \\
\hline
\multicolumn{4}{l}{\textbf{TACO Improvement over Qiskit}} & \textit{1.26$\times$} & \textit{490$\times$} & \textit{352$\times$} \\\hline
\end{tabular}
\label{tab:qiskit_compare}
\end{table}

We compare \ours against Qiskit for Clifford+T transpilation. We focus on three key metrics: the number of unitaries requiring synthesis, the number of T gates, and transpilation time. Of the eight benchmark circuits, the first four (qft, qpe, ising, and w\_state) include arbitrary rotation gates that require synthesis, while the others can be trivially decomposed into Clifford+T gates. Therefore, we only include comparisons for the non-trivial benchmarks. Results are shown in Table~\ref{tab:qiskit_compare}. \ours reduces the number of unitaries requiring synthesis by 1.26$\times$ on average, even compared to Qiskit with O3 optimization. \ours achieves 490$\times$ fewer T gates on average, which is attributed to both fewer unitaries requiring synthesis and the use of a more efficient synthesis algorithm. More importantly, \ours achieves these superior results with an average 352$\times$ faster transpilation time.

\subsection{Applicability to Prior Clifford+T Synthesis}
\label{sec:compare_prior_synthesis}

Synthesizing algorithmic circuits into fault-tolerant Clifford+T circuits is an important step in the FTQC toolchain, and several prior works have made significant progress on this problem, including Synthetiq~\cite{paradis2024synthetiq} and TRASYN~\cite{hao2025reducing}. These approaches are orthogonal to \ours: while they synthesize algorithmic circuits into Clifford+T form, \ours operates \emph{after} synthesis and further reduces the remaining Clifford overhead in the resulting circuits. One possible concern, however, is that synthesized Clifford+T circuits may exhibit less regular structure, potentially making the Clifford-reduction techniques used by \ours more challenging to apply. To evaluate the robustness of \ours in this setting, we collect the synthesized benchmark circuits reported in the original Synthetiq and TRASYN papers and apply \ours to them. For each comparison, we ask two questions: (1) whether Clifford gates still dominate the total gate count after synthesis, and (2) how much of those Clifford gates can be eliminated by \ours.

\begin{figure}[h!]
    \centering
    \includegraphics[width=\columnwidth]{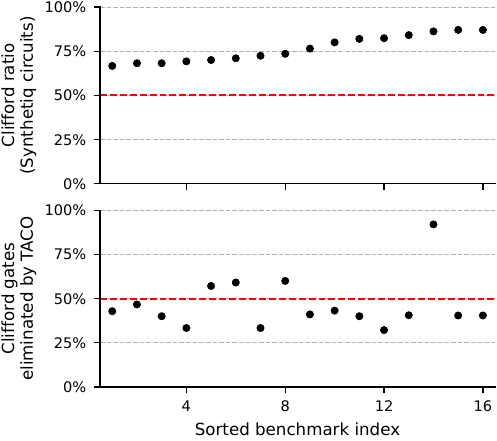}
    \caption{Across 16 synthesized Clifford+T circuits from Synthetiq, whose sizes range from 10 to 68 total Clifford+T gates, Clifford gates remain the majority after synthesis and \ours still eliminates a substantial fraction of them. \textbf{Top:} Clifford-gate ratio in each synthesized circuit. \textbf{Bottom:} fraction of Clifford gates eliminated by \ours on the same circuit. Benchmarks are sorted in ascending order of Clifford ratio in the top panel, and the bottom panel follows the same order for direct correspondence.}
    \label{fig:synthetiq_compare}
\end{figure}

Figure~\ref{fig:synthetiq_compare} shows the results on the 16 synthesized Clifford+T circuits reported by Synthetiq. Although these benchmarks are fairly small, ranging from 10 to 68 total Clifford+T gates, Clifford gates still dominate every synthesized circuit, with a minimum Clifford ratio of 66.7\% and a median of 76.5\%. This confirms that even after synthesis, Clifford gates remain the majority component. At the same time, because Synthetiq is designed to synthesize individual operations, the resulting circuits are relatively small, leaving less room for further optimization. Even in this constrained setting, \ours still removes a substantial fraction of the Clifford gates, achieving 49.1\% reduction on average, 41.0\% at the median, and up to 92.9\% in the best case.

\begin{figure}[h!]
    \centering
    \includegraphics[width=\columnwidth]{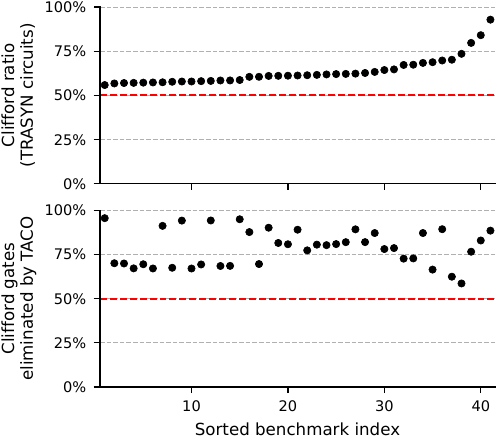}
    \caption{Across 41 synthesized Clifford+T circuits from TRASYN, whose sizes range from 28 to 26,215 total Clifford+T gates, Clifford gates remain the majority after synthesis and \ours still eliminates a substantial fraction of them. \textbf{Top:} Clifford-gate ratio in each synthesized circuit. \textbf{Bottom:} fraction of Clifford gates eliminated by \ours on the same circuit. Benchmarks are sorted in ascending order of Clifford ratio in the top panel, and the bottom panel follows the same order for direct correspondence.}
    \label{fig:trasyn_compare}
\end{figure}

Figure~\ref{fig:trasyn_compare} shows the corresponding results on the 41 synthesized Clifford+T circuits from TRASYN. In contrast to Synthetiq, TRASYN performs synthesis at the circuit level and therefore produces benchmarks that are much larger overall, ranging from 28 to 26,215 total Clifford+T gates. Nevertheless, Clifford gates still remain the majority in all synthesized circuits, with a minimum Clifford ratio of 55.8\% and a median of 61.2\%. More importantly, \ours continues to achieve strong reductions in this larger-scale setting, eliminating 78.7\% of Clifford gates on average and 80.2\% at the median, with a maximum reduction of 95.5\%. This reduction is much closer to the savings achieved on our own benchmark suite, indicating that \ours remains highly effective even when applied to optimized Clifford+T circuits produced by state-of-the-art circuit synthesis works.

Overall, these results show that prior synthesis and \ours are complementary rather than competing. Existing tools such as Synthetiq and TRASYN reduce the cost of converting algorithmic descriptions into Clifford+T form, while \ours can be applied afterward to further reduce the Clifford overhead that still dominates the synthesized circuits.

\section{Related Work}
\subsection{Pauli-Based Computation}
\label{sec:related_pbc}

PBC is a framework for analyzing and optimizing quantum circuits by representing quantum gates as Pauli rotation operations and studying their commutation relationships through Pauli strings. Different works have employed this framework for various purposes: Litinski used PBC to eliminate Clifford gates in fault-tolerant architectures~\cite{litinski2019game}, while others have applied PBC to reduce resources in NISQ circuits~\cite{Peres2023quantumcircuit}, to extend the formalism to higher-dimensional systems~\cite{Peres2023PBC}, or to design PBC-inspired optimization methods~\cite{Paykin2023PCOAST}. However, these non-FTQC optimizations are not applicable to FTQC, as they introduce additional operations that cannot be executed fault-tolerantly. Therefore, the only relevant prior work to our setting is Litinski’s PBC-based approach for eliminating Clifford gates. While PBC is effective when T-magic-state generation is the bottleneck~\cite{litinski2019game}, recent advances~\cite{litinski2019magic, gidney2024magic} have lowered T-state costs, shifting the critical challenge to circuit parallelism. Thus, \ours's ability to eliminate Clifford overhead while preserving parallelism is essential for realizing the quantum speedups envisioned in prior work~\cite{gidney2021factor}.

\subsection{QEC Codes and Optimization}

This paper focuses on patch-based surface code with lattice surgery. Other QEC constructions include defect-based~\cite{fowler2012surface, horsman2012surface, litinski2019game} and twist-based~\cite{Bombin2010} surface code. Overlapping multiple surface code patches on neutral atom devices was also explored~\cite{viszlai2025interleaved}. Beyond surface codes, quantum stabilizer codes like Shor's~\cite{shor1995scheme, Bacon2006} and Steane's~\cite{Steane1996} use stabilizer groups to define protected logical subspaces. Color codes~\cite{Bombin2006} support transversal Clifford operations and have been experimentally shown~\cite{Nigg2014, Ryan-Anderson2021, Hilder2022, bluvstein2024logical}. Quantum LDPC codes~\cite{Bravyi2024, Lawrence2022, Breuckmann2021} feature long-range checks and higher code rates. Fault-tolerant non-Clifford operations require magic states. Various distillation protocols have been proposed to prepare high-fidelity magic states using Reed-Muller code~\cite{bravyi2005universal, Haah2018codesprotocols}, block codes~\cite{bravyi2012magic, jones2013multilevel, fowler2013}, and recent optimizations~\cite{Campbell2018magicstateparity, litinski2019magic, gidney2021factor, gidney2024magic}. QEC decoding uses syndrome measurements to identify errors through lookup tables~\cite{Tomita2014, Das2022}, MWPM~\cite{dennis2002topological, Smith2023, higgott2023sparseblossom, wu2023fusionblossom, Vittal2023, alavisamani2024,deMartiiOlius2024decodinga}, or machine learning~\cite{Chamberland2023, Andreasson2019quantum, Baireuther2019}. System optimizations for synchronizing lattice surgery~\cite{maurya2025synchronization} and speculative decoding~\cite{viszlai2025swiper} have also been explored.

\subsection{General Clifford Circuit Optimization}
\label{sec:related_general_optimization}

Several works have focused on optimizing circuits containing Clifford gates. Maslov et al.\cite{maslov2008quantum} and Bravyi et al. \cite{bravyi2021clifford} employ template-based circuit optimization techniques, which attempt to replace sequences of Clifford gates with more efficient alternatives. However, in Clifford+T circuits, the consecutive Clifford-only gates sandwiched between T gates are typically very short (usually fewer than 3 gates), making template-matching techniques ineffective. In contrast, \ours studies how to eliminate Clifford gates in the presence of non-Clifford gates. Liu et al.~\cite{liu2024qucloud+} proposed absorbing Clifford gates into neighboring gates, resulting in new composite gates. While such gates might be executable in NISQ devices, they cannot be fault-tolerantly executed under QEC schemes.

\section{Conclusion}
 Optimizing Clifford gates remains a key challenge for enabling practical FTQC. We present \ours, a Transpiler-Architecture Co-design Optimization framework that closes this gap, achieving an average 91.2\% reduction in Clifford gates across a range of quantum circuits while fully preserving gate parallelism. \ours delivers up to 21.9$\times$ speedup over PBC, with a geometric mean speedup of 4.4$\times$ across benchmarks. Our co-designed FTQC architecture further boosts efficiency by exploiting the locality of optimized rotation sequences, enabling one logical gate per QEC cycle with just $1.5n + 4$ logical qubit tiles. Together, these cross-stack optimizations at both circuit and architectural levels mark a significant advance toward efficient, practical FTQC.

\section*{Acknowledgment}
This material is based upon work supported by the U.S. Department of Energy, Office of Science, National Quantum Information Science Research Centers, Co-design Center for Quantum Advantage (C2QA) under contract number DE-SC0012704 (Basic Energy Sciences, PNNL FWP 76274). This research was also supported in part by the National Research Council (NRC) Canada grants AQC 003 and AQC 213, as well as the Natural Sciences and Engineering Research Council of Canada (NSERC) [funding number RGPIN-2019-05059]. This research used resources and associated infrastructure support of the Oak Ridge Leadership Computing Facility, which is a DOE Office of Science User Facility supported under Contract DE-AC05-00OR22725. This research used resources of the National Energy Research Scientific Computing Center (NERSC), a U.S. Department of Energy Office of Science User Facility located at Lawrence Berkeley National Laboratory, operated under Contract No. DE-AC02-05CH11231.

\section*{Appendix: Artifact Evaluation}

\subsection{Abstract}

This artifact provides the Zenodo reproducibility package for the \textbf{TACO} framework and the experimental workflow used in the paper. It contains the benchmark datasets, figure-generation scripts, and a self-contained copy of the NWQEC codebase with TACO functionality integrated. The artifact is available at \url{https://doi.org/10.5281/zenodo.19449157}, and NWQEC is open-sourced on GitHub at \url{https://github.com/pnnl/nwqec}. The package reproduces key results on Clifford reduction and gate parallelism via automated command-line workflows that generate intermediate CSV files and final PDF figures. All experiments run on a standard CPU machine without specialized hardware and complete within tens of minutes. Detailed build and execution instructions are provided in the artifact \texttt{README}.

\subsection{Artifact check-list (meta-information)}

{\small
\begin{itemize}
  \item {\bf Algorithm: } Clifford+T and Pauli-based computation (PBC) FTQC transpilation;
Clifford-reduction analysis; gate-parallelism analysis
  \item {\bf Program: } NWQEC (\texttt{nwqec-cli}), Bash workflows, Python plotting/analysis scripts
  \item {\bf Compilation: } CMake + C++17
  \item {\bf Transformations: } N/A
  \item {\bf Binary: } \texttt{nwqec-cli}
  \item {\bf Data set: } QASM benchmark circuits (paper benchmark set + prior-work benchmark sets)
  \item {\bf Run-time environment: } Linux or macOS shell environment with Python 3
  \item {\bf Hardware: } CPU-only
  \item {\bf Execution: } Scripted command-line workflows via top-level Bash scripts
  \item {\bf Metrics: } Clifford ratio, Clifford reduction ratio, gate parallelism, and operation-weight distributions
  \item {\bf Output: } CSV files in \texttt{results/} and PDF figures in \texttt{figures/}
  \item {\bf Experiments: } Fig.~\ref{fig:op_weights}, Fig.~\ref{fig:gate_parallelism}, Fig.~\ref{fig:clifford_reductions}, Fig.~\ref{fig:clifford_reduction_scaling}, Fig.~\ref{fig:synthetiq_compare}, Fig.~\ref{fig:trasyn_compare}
  \item {\bf How much disk space required (approximately)?: } $<$ 100 MB
  \item {\bf How much time is needed to prepare workflow (approximately)?: } 5–10 minutes (build + environment setup)
  \item {\bf How much time is needed to complete experiments (approximately)?: } 10–20 minutes (Fig.~20 is the longest)
  \item {\bf Publicly available?: } yes
  \item {\bf Code licenses (if publicly available)?: } MIT
  \item {\bf Data licenses (if publicly available)?: } MIT
  \item {\bf Workflow automation framework used?: } Bash scripts
  \item {\bf Archived (provide DOI)?: } 10.5281/zenodo.19449157
\end{itemize}
}

\subsection{Description}

\subsubsection{How to access}

The artifact can be downloaded from the DOI link:
\url{https://doi.org/10.5281/zenodo.19449157}

\subsubsection{Hardware dependencies}

No specialized hardware is required. All experiments run on a standard
CPU-based workstation or laptop.

\subsubsection{Software dependencies}

\begin{itemize}
  \item CMake and a C++17 compiler
    \item GMP and MPFR, used by the synthesis backend. On macOS and Linux, these dependencies need not be installed manually: NWQEC automatically uses precompiled binaries, downloading them if they are not already present
  \item Python 3 with plotting/data packages used by scripts
\end{itemize}

This artifact has been tested on macOS and Linux.

\subsubsection{Data sets}

Benchmark circuits are included:
\begin{itemize}
  \item \texttt{benchmarks/}: benchmark circuits used in the main paper experiments
  \item \texttt{synthetiq/}: circuits generated by the Synthetiq benchmark suite
  \item \texttt{trasyn/}: circuits generated by the TRASYN benchmark suite
\end{itemize}

\subsubsection{Models}

N/A.

\subsection{Installation}

Build NWQEC from the artifact root directory:

{\small
\begin{verbatim}
cmake -S . -B build -DCMAKE_BUILD_TYPE=Release
cmake --build build -j
\end{verbatim}
}
\subsection{Experiment workflow}

The artifact provides automated scripts to reproduce the experimental
results reported in the paper.

For convenience, the full workflow can be executed with a single script:

{\small
\begin{verbatim}
./run_all.sh
\end{verbatim}
}

This script performs a fresh build of the NWQEC binary, runs all
experiments on the benchmark circuits, collects metrics into CSV files
under \texttt{results/}, and generates the corresponding PDF figures
under \texttt{figures/}. 

Individual figures can also be reproduced using dedicated top-level
scripts (one per figure). For example:

{\small
\begin{verbatim}
./plot_fig_5.sh
\end{verbatim}
}

Each figure script \textbf{both runs the necessary experiment and generates
the final plotted figure}. If the corresponding CSV results already
exist, the experiment is skipped to save time, and only the figure is
generated. Passing the \texttt{--force-collect} flag forces the
script to re-run the experiment and regenerate the data.

Detailed usage instructions and script options are provided in the
artifact \texttt{README}.

\subsection{Evaluation and expected results}

The artifact should regenerate the following figure PDFs in \texttt{figures/}:
\begin{itemize}
  \item Fig.~5: QFT operation-weight distribution
  \item Fig.~14: QFT gate parallelism
  \item Fig.~17: Clifford reduction on benchmark set
  \item Fig.~20: large-QFT reduction trend
  \item Fig.~21: Synthetiq comparison
  \item Fig.~22: TRASYN comparison
\end{itemize}

Numeric outputs are generated as CSV files in \texttt{results/}. The figures should match trends and relative values reported in the paper. Minor numerical or formatting differences may occur due to environment or library variations.

\subsection{Experiment customization}

Scripts support standard environment-variable overrides for the Python interpreter and the NWQEC binary path.
The README documents available script flags for forced regeneration and figure-specific customization.

\subsection{Notes}

The artifact is designed to reproduce the reported trends and relative improvements rather than to serve as a complete benchmark suite. Generated outputs may differ slightly across environments due to library versions, compiler settings, or plotting backends.

\subsection{Methodology}

Submission, reviewing, and badging methodology:

\begin{itemize}
  \item \url{https://www.acm.org/publications/policies/artifact-review-and-badging-current}
  \item \url{https://cTuning.org/ae}
\end{itemize}

\balance
\bibliographystyle{IEEEtranS}
\bibliography{references}

\end{document}